\documentclass{emulateapj}
\usepackage{graphicx}
\usepackage{epsfig}
\usepackage{natbib}
\usepackage{multirow}
\usepackage{amsmath,amssymb}
\usepackage{rotating}

\begin{document}

\title{Submillimeter sources behind the massive lensing clusters A370 and A2390}
\author{Chian-Chou Chen\altaffilmark{1}, Lennox L. Cowie\altaffilmark{1}, Wei-Hao Wang\altaffilmark{2}, Amy J. Barger\altaffilmark{1,3,4}, Jonathan P. Williams\altaffilmark{1}}
\altaffiltext{1}{Institute For Astronomy, University of Hawaii, 2680 Woodlawn Drive, Honolulu, HI 96822, USA}
\altaffiltext{2}{Academia Sinica Institute of Astronomy and Astrophysics, P.O. Box 23-141, Taipei 10617, Taiwan}
\altaffiltext{3}{Department of Astronomy, University of Wisconsin-Madison, 475 North Charter Street, Madison, WI 53706.}
\altaffiltext{4}{Department of Physics and Astronomy, University of Hawaii, 2505 Correa Road, Honolulu, HI 96822.}
\subjectheadings{cosmology: observations  |  galaxies: evolution  |  galaxies: formation  |  galaxies: high-redshift | submillimeter: galaxies}

\begin{abstract}
We report 850~$\mu$m Submillimeter Array (SMA) observations of four gravitationally lensed submillimeter galaxies (SMGs), A370-2, A2390-1, A2390-3 and A2390-4, which were originally discovered with the Submillimeter Common-User Bolometer Array (SCUBA). Our SMA detection of A370-2 with a submillimeter flux of 7.95 $\pm$ 0.60~mJy unambiguously identifies the counterparts to this source at optical and radio wavelengths. A2390-1 is an ultraluminous infrared galaxies with a submillimeter flux of 5.55 $\pm$ 0.92~mJy and a redshift of 1.8 $\pm$ 0.2 computed from submillimeter/radio flux ratio analysis. We resolve A2390-3 into two components, A2390-3a and A2390-3b, with fluxes of 3.15 $\pm$ 0.63~mJy and 1.92 $\pm$ 0.60~mJy, respectively. The structure of the system could be consistent with morphological distortion by gravitational lensing. The lack of counterparts in the optical and infrared indicates a heavily dust-enshrouded nature, and a non-detection in the radio implies that these two sources probably lie at $z > 4.7$, which would make them among the most distant SMGs known to date. Our non-detection of A2390-4 suggests either that there are multiple fainter submillimeter sources within the SCUBA beam or that the SCUBA detection may have been false. Our precise positions allow us to determine accurate amplifications and fluxes for all our detected sources. Our new results give a shallower power-law fit (-1.10) to the faint-end 850~$\mu$m cumulative number counts than previous work. We emphasize the need for high-resolution observations of single-dish detected SMGs in order to measure accurately the faint end of the 850~$\mu$m counts.
\end{abstract}

\section{Introduction}
During the past decade, deep blank-field submillimeter/millimeter surveys from the Submillimeter Common-User Bolometer Array (SCUBA) on the James Clerk Maxwell Telescope\footnote[5]{The JCMT is operated by the Joint Astronomy Centre on behalf of the Science and Technology Facilities Council in the United Kingdom, the Netherlands Organization for Scientific Research, and the National Research Council of Canada} (JCMT) and the Max Planck Millimeter Bolometer (MAMBO) array on the IRAM 30~meter telescope have resolved sources brighter than $\sim$ 2~mJy that account for $\sim$ 20\%--30\% of the 850~$\mu$m extragalactic background light (EBL) \citep{Barger:1999p6485,Eales:2003p6489,Scott:2002p6539,Webb:2003p6591,Borys:2003p6612,Greve:2004p6618,Wang:2004p2270,Coppin:2006p9123}. The rest of the 850~$\mu$m EBL can be accounted for by sources over the 0.3--2~mJy range measured using gravitational lensing \citep{Blain:1999p7279, Cowie:2002p2075, Knudsen:2008p3824}.

However, further studies on individual submillimeter galaxies are hampered by the relatively poor resolution of single-dish submillimeter telescopes (e.g., $15''$ FWHM at 850~$\mu$m on SCUBA). This causes a correspondingly large uncertainty in the source positions and confusion in identifying counterparts to the submillimeter sources at other wavelengths.  It also makes the gravitational lensing uncertain, particularly for sources with very high amplification. Techniques have been developed to work around this problem: The empirical correlation between non-thermal radio emission and thermal dust emission \citep{Condon:1992p6652}, together with the high astrometric precision of radio interferometers, has been widely used as a tool to study SMGs. \citet{Chapman:2005p5778} determined the redshifts of 73 radio-identified SMGs using Keck spectroscopy, finding a median redshift of $\sim$ 2.2. It has also been found that SMGs appear to have similar properties to the local ultraluminous infrared galaxies with far-infrared (FIR) luminosities of $\sim$ 10$^{12}$--10$^{13}$ \textit{$L_\odot$} \citep{Sanders:1996p6419}. The lack of strong X-ray emission from submillimeter sources (\textit{L$_X$/L$_{FIR}$} = 0.004) suggests that most of the FIR/submillimeter flux is dominated by dust-reradiated emission from intense star formation rather than AGN output \citep{Alexander:2005p6453}, which makes studying the FIR/submillimeter sources a path to robustly determining the dust-obscured star formation history.

There are several drawbacks to using radio emission to identify the SMGs. Firstly, with the current sensitivity (5 $\sigma \sim 20~\mu$Jy at 20~cm) of radio interferometers \citep{Morrison:2010p7451}, the radio-identified SMGs are mostly bright ($\gg$ 2~mJy at 850 $\mu$m), so their properties may not be representative of the submillimeter population as a whole. Secondly, the radio flux drops at high redshift due to the positive \textit{K}-correction, whereas the submillimeter remains almost invariant over the redshift range $z\sim 1 - 8$ \citep{Blain:2002p8120}. Thus, the radio-identification technique is biased against high-redshift sources. Lastly, there are cases where multiple submillimeter sources are located within the beam of a single-dish submillimeter telescope \citep{Wang:2011p9293}, so a radio source within the beam may not be the correct or only counterpart.

\begin{table*}
 \begin{center}
 \caption{SMA observations}
 \scalebox{1}{
 \begin{tabular}{cccccccc}
 \hline
 \hline
 Target & R.A. (2000)  & Decl. (2000) & SCUBA S$_{850~{\rm \mu m}}^a$  & SMA S$_{850~{\rm \mu m}}$ & Amplification$^a$  & Theoretical Noise (1 $		 \sigma$)$^b$  &  Resolution$^b$ \\
 Source & (h m s) & (d m s) &(mJy)&(mJy)&(min,max)& (mJy beam$^{-1}$) &  \\
 \hline
 A370-2 & 02 39 56.63 & -1 34 27.0 & 6.68 $\pm$ 0.58  & 7.95 $\pm$ 0.60 & 2.3 & 0.57  &  $2\farcs4 \times 2\farcs3$   \\
 A2390-1 & 21 53 33.58 & 17 42 42.3 & 7.57 $\pm$ 0.93 & 5.55 $\pm$ 0.92 & 1.9 (1.8,2.0) & 0.77 & $2\farcs3 \times 1\farcs$9\\ 
 A2390-3 & 21 53 35.48 & 17 41 09.3 & 3.24 $\pm$ 0.78 & 5.07 $\pm$ 0.87 & 52 (0.6,52) & 0.55  &  $2\farcs0 \times 1\farcs6$  \\
 A2390-4 & 21 53 38.21 & 17 41 52.3 & 2.64 $\pm$ 0.72 & $<$ 0.96$^c$ & 11 ($>$ 6.7)  &  0.32  &  $2\farcs2 \times 1\farcs8$ \\
 \hline
 &&&&&&\\
 \multicolumn{8}{l}{$^{a}$ \citet{Cowie:2002p2075} SCUBA 850~$\mu$m fluxes and source amplifications, which were determined using the SCUBA positions and} \\ 
 \multicolumn{8}{l}{ LENSTOOL.} \\
 \multicolumn{7}{l}{$^{b}$ Obtained from natural weighted baselines. The resolution is the FWHM of the synthesized beam.} \\
 \multicolumn{7}{l}{$^{c}$ The 3~$\sigma$ upper limit for a point source.}
 \end{tabular}
 }
 \label{source}
 \end{center}
\end{table*}

In the last few years our ability to address these issues has greatly improved with the advent of telescopes like the Submillimeter Array\footnote[6]{The Submillimeter Array is a joint project between the Smithsonian Astrophysical Observatory and the Academia Sinica Institute of Astronomy and Astrophysics and is funded by the Smithsonian Institution and the Academia Sinica.} (SMA; \citealt{Ho:2004p8376}). The SMA provides imaging in the submillimeter regime with arcsecond resolution \citep{Iono:2006p6985,Wang:2007p6971,Younger:2007p6982,Younger:2008p8372,Cowie:2009p6978,Aravena:2010p8370}. Some distances to submillimeter sources have been found from spectroscopic redshifts \citep{Capak:2008p8406,Coppin:2009p8366} and found or confirmed from CO line searches \citep{Schinnerer:2008p9300,Daddi:2009p9295,Daddi:2009p9297,Coppin:2010p8367,Riechers:2010p9307}. SMA observations have shown that not all SMGs have strong radio counterparts \citep{Younger:2007p6982,Younger:2008p8372,Cowie:2009p6978}. This again illustrates the problems with the radio identification technique. However, most of the sources studied so far have fluxes around 10~mJy, and they constitute only a small fraction of the submillimeter EBL ($\sim$ 2\%, according to the counts in Coppin et al.\ 2006). The nature of more typical SMGs with submillimeter fluxes $<2$~mJy is still an open question. 

With the improved sensitivity from the recently upgraded double bandwidth (4~GHz) on the SMA, we are now able to reach a sub-mJy level of sensitivity and probe relatively submillimeter-dim sources. In order to sample very faint sources we have used this capability to observe sources in the gravitationally lensed regions of two massive clusters, A370 and A2390 \citep{Cowie:2002p2075}, which enables us to take advantage of the lensing amplifications. We list the basic information on our target sources in Table \ref{source}, where we follow the index sequence from the catalog of \citet{Cowie:2002p2075}. We give the observational details and discuss our reductions in Section~2. In Section~3 we report on our SMA observations of one radio-bright source, A370-2, and two radio-dim sources, A2390-3 and A2390-4.  We show unambiguously whether there are counterparts to these sources in the optical, infrared, or radio. In Section~4 we discuss the implications of our new measurements on the faint-end 850~$\mu$m galaxy number counts. We summarize our results in Section~5. We assume the WMAP (\textit{Wilkinson Microwave Anisotropy Probe}) cosmology throughout: \textit{H}$_0$ = 70.5 km s$^{-1}$ Mpc$^{-1}$, $\Omega_M$ = 0.27, $\Omega$$_\Lambda$ = 0.73 \citep{Komatsu:2009p8122}.

\begin{figure*}
 \begin{center}
    \leavevmode
      \includegraphics[width=1 \textwidth] {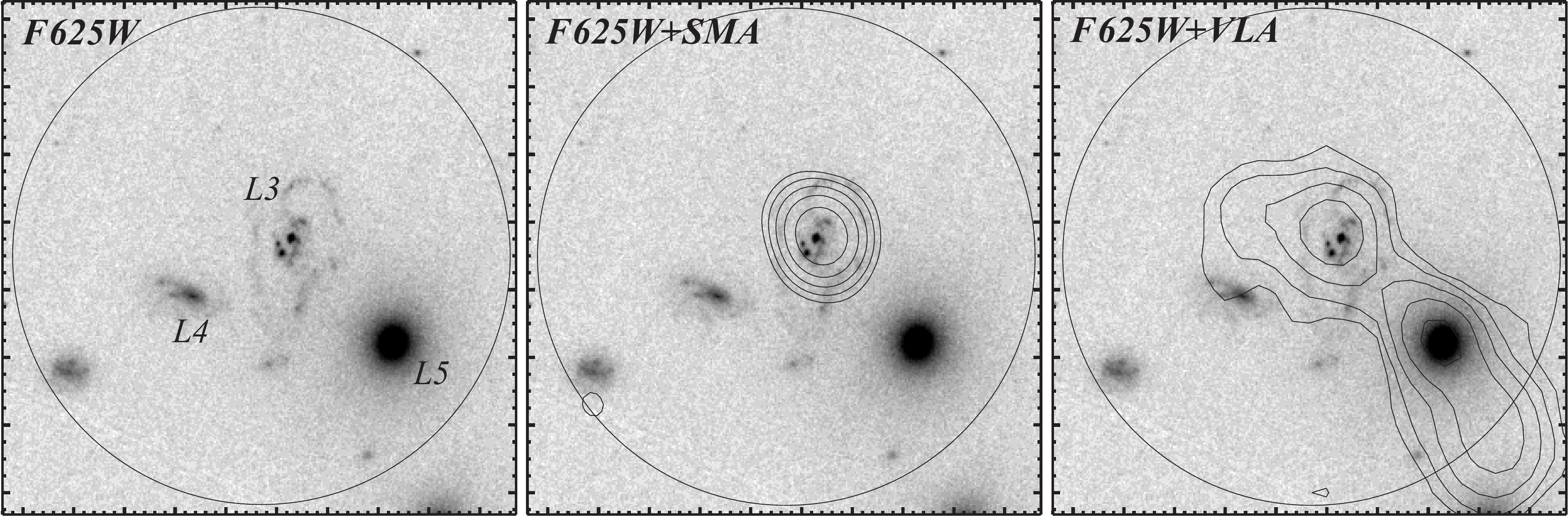}
       \caption{Multiwavelength images of the ring galaxy A370-2 and its surroundings: \textit{Left}: \textit{HST} ACS F625W image -- \textit{Center}: SMA 850~$\mu$m contours overplotted on the F625W image -- \textit{Right}: 20~cm VLA contours overplotted on the F625W image. Note that all three images are centered at the position given in Table~\ref{source}. The big black circle in each image shows the SCUBA beam size (15$''$ $\times$ 15$''$). North is up and East is to the left. }
     \label{a370}
  \end{center}
\end{figure*}

\section{Observations and Data Reduction}

Seven 345~GHz SMA tracks of A370-2, four tracks of A2390-1, three tracks of A2390-3 and three tracks of A2390-4 were taken between 2007 and 2010. We used the subcompact configuration (16.5 -- 32~meter baselines) on two tracks of A370-2 and the compact configuration (20 -- 70~meter baselines) on all other tracks in order to obtain a better signal to noise ratio. The shape of the bandpass was measured by observing two quasars 3C454.3 and 3C111. The sources 0238+166, 0132--169, 2148+069 and 2232+117 were observed for time dependent gain calibration, and the planets Uranus, Ceres and Callisto were used as flux calibrators. We used the data reduction package MIR to calibrate the visibilities, and we produced the images using the MIRIAD routines \citep{Sault:1995p6974}. 

To minimize the noise level we combined all the tracks for each source and made the final continuum images with naturally weighted baselines. The cleaned images (the outcome of the CLEAN algorithm) of the calibrators from each and every track show only one cleaned component at the phase center of the calibrators. After performing RESTOR to convolve the cleaned component with the point spread function,  every calibrator appeared as a point source, which is an indication of good phase calibration. In order to prevent over-cleaning of the dirty maps, we also made sure that the r.m.s of the cleaned maps, excluding the detected sources, agreed with the theoretical noise. One should keep in mind that there is a typical $\sim10\%$ error introduced by the flux calibration, on top of the image r.m.s noise for flux measurements. 

The theoretical r.m.s noise at the phase center and the resolution of each of the final images of the four SMGs are given in Table \ref{source}. Note that the better weather conditions for A2390-3 and the double bandwidth (4~GHz) used in the last track of the A2390-3 data make the sensitivity comparable to A370-2, even though there were four fewer tracks of data for A2390-3 than for A370-2. The sensitivity of the image of A2390-1 is not as good as others due to bad weather conditions during the first two tracks of data. For A2390-4 all three tracks of data are double bandwidth, so a better sensitivity was obtained.

\section{Results}

We have a strong detection ($>$ 5~$\sigma$) toward A370-2 and A2390-1, respectively, and two significant detections (both $>$ 3 $\sigma$) toward A2390-3. However, we found no sources at the A2390-4 position, even with these deepest SMA observations toward a single SMG. It is possible that there is more than one source contributing to the submillimeter emission in this region, and that once we are able to resolve it into individual sources, their fluxes will be found to be below the current detection limit, as has been the case for several submillimeter sources in the GOODS-N \citep{Wang:2011p9293}. Alternatively, the SCUBA detection may have been false. Again, this is one of the main reasons to use high-resolution telescopes to observe SCUBA sources. We discuss the A370-2, A2390-1 and A2390-3 detections below.

\subsection{A370-2}

A370-2, which is also referred to as SMM J02399--0134 \citep{Barger:1999p6801,Smail:2000p6377} or SMM J02396-0134 \citep{Greve:2005p6788,Wu:2009p6792}, has three optical sources within the SCUBA beam (Figure \ref{a370}). The galaxy L3 has a distorted ring morphology. Its formation mechanism has been suggested to be either the dynamical axial penetration of a smaller companion galaxy into a disk galaxy or a pair of superbubbles driven by the intense starburst at the central region \citep{Taniguchi:2001p6381}. A spectroscopic redshift of $\sim$1.06 was independently obtained by two groups using LRIS on Keck II and OSIS-V on the Canada-France-Hawaii Telescope (CFHT), respectively, and the identification of the high-ionization lines of [Ne V] indicates that L3 is likely to be a Seyfert galaxy \citep{Barger:1999p6801,Soucail:1999p6372}. Located beyond the cluster A370, the signal from L3 is amplified by a factor of 2.3 due to gravitational lensing \citep{Cowie:2002p2075}. A strong CO molecular detection toward L3 was reported by \citet{Greve:2005p6788} using the IRAM Interferometer. L4 is at $z=0.42$ and appears to be a background normal galaxy. The bright passive elliptical galaxy L5 is one of the cluster members at $z = 0.37$ (\#32 in \citealt{Mellier:1988p6845}). 

The evidence discussed above implies that L3 is the only active source in this region, and it has been argued that L3 is the likely counterpart of the SMG \citep{Barger:1999p6801}. Our SMA data confirm this identification and clearly demonstrate that the submillimeter emission is from the center of galaxy L3 (Figure 1).  A370-2 is detected both in our dirty and CLEANed images. The point-source fit routine IMFIT in MIRIAD with a 10$''$ box centered at the peak position produces a flux of 7.95~mJy and also yields an r.m.s for the residual image of 0.60~mJy/beam. Note the flux measurements throughout this paper are all primary beam corrected. The fitted peak position is at $\alpha$(J2000.0) = 2$^h$39$^m$56.55$^s$, $\delta$(J2000.0) = -1$^\circ$34$^\prime$26.54$^{\prime\prime}$ with an error of $0\farcs15$. 
 
The total submm flux of 7.95 $\pm$ 0.60~mJy agrees with the SCUBA flux of 6.68 $\pm$ 0.58~mJy from \citet{Cowie:2002p2075} (Table \ref{source}). We use the spectral energy distribution (SED) model from  \citet{Barger:2000p2144} based on an Arp~220 template with an assumed dust temperature and extinction coefficient ($T_d$ = 47~K, $\beta$ = 1) to compute the infrared luminosity. The SED in the infrared and radio regime is scaled to fit our 850~$\mu$m flux, and the infrared luminosity can simply be calculated from the area under the SED curve between 8~$\mu$m and 1000~$\mu$m. With the adopted redshift of 1.06, the de-lensed flux of 3.46~mJy corresponds to an infrared luminosity of $\sim 2.5 \times 10^{12}~L_\odot$, which is typical of ultraluminous infrared galaxies (ULIRG, $L_{IR} > 10^{12}~L_\odot$).

A strong 1.4~GHz (20~cm) detection of the source was reported with $\sim 5\farcs0$ resolution using the Very Large Array \citep{Smail:2000p6377}. The emission peaks at a location close to L5, and the morphology appears unresolved and elongated from north-east to south-west. In order to resolve this region, we made use of the higher resolution ($1\farcs68 \times 1\farcs49$) VLA archival data taken on 1999 August at the same frequency toward this region in the most extended A configuration. The data clearly show two radio sources at the locations of the optical sources L3 and L5 (Figure 1). Note that without our SMA observations, it would still be ambiguous which radio source is the counterpart of A370-2 based on the correlation between submillimeter and radio emission. 

\begin{figure}[h]
 \begin{center}
    \leavevmode
      \includegraphics[scale=0.48]{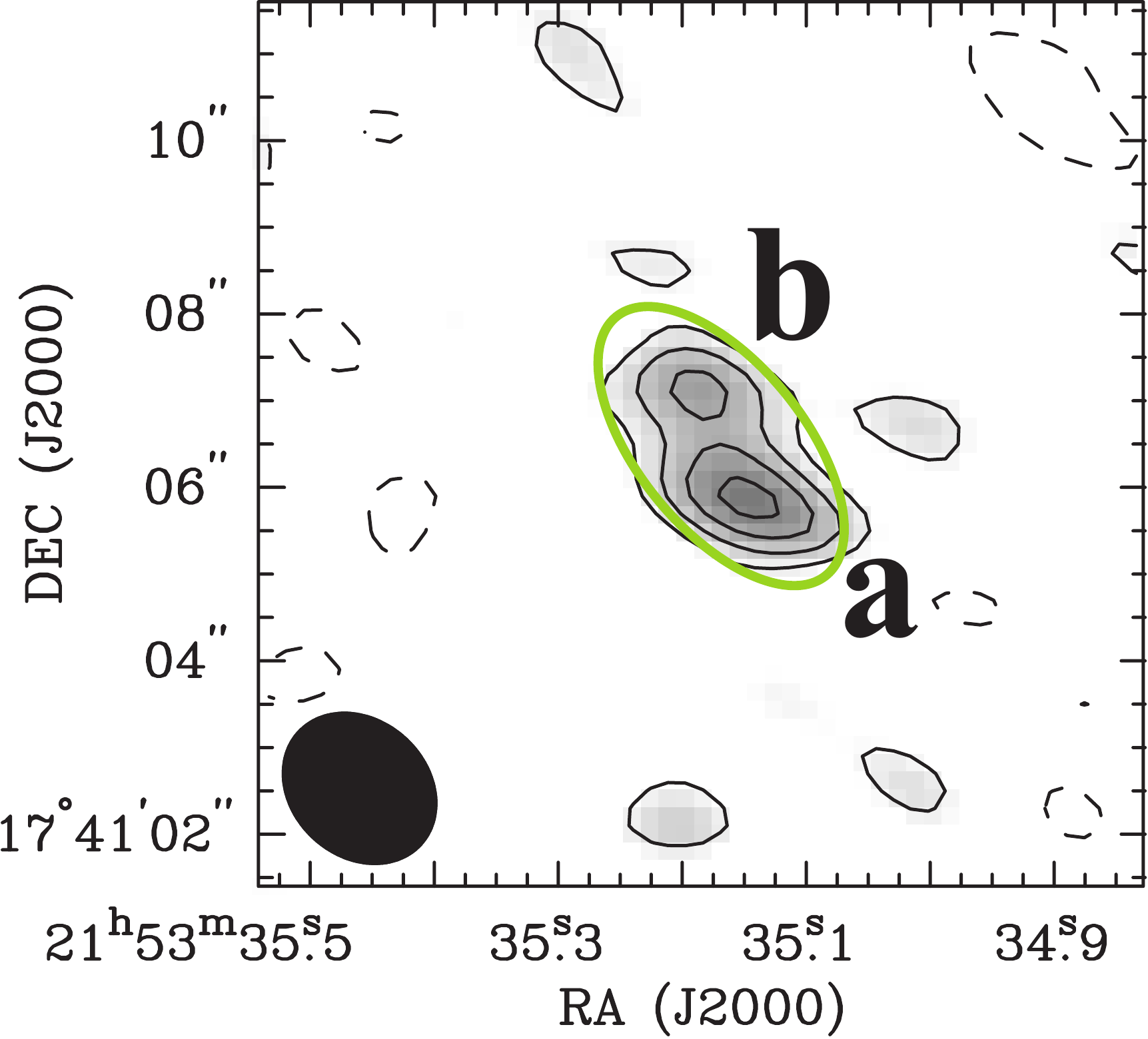}
       \caption{A2390-3 SMA image showing the two submillimeter detections marked by the letters a and b. The black contours have levels of $-2, 2, 3, 4, 5~\sigma$, where 1~$\sigma$ is 0.58~mJy beam$^{-1}$. The gray scale image shows the linear scale range from 1.0 to 5.0~mJy beam$^{-1}$. A $1\farcs9 \times 1\farcs6$ beam is presented in the lower left corner. The green ellipse is the outcome of gravitational lensing, assuming a 1$''$ radius source in the middle of two sources on the source plane. }
     \label{a2390sma}
  \end{center}
\end{figure}

The radio and submillimeter peak positions at the L3 location are still offset by $0\farcs6$, which may be caused by the contamination of the jet structure extending from L5. The radio flux from L3, 0.573~mJy, is estimated from a $2\farcs5$ radius circular aperture covering most of the radio flux from L3. We ran a Monte Carlo simulation with the same aperture on the VLA image, and this gave a noise level of 0.011~mJy. Together with the submillimeter flux, our results show a $S_{850~\mu {\rm m}}$ to $S_{\rm 1.4~GHz}$ ratio of 12.6 -- 15.2. For sources lying at $z < 3$ in the Rayleigh-Jeans long-wavelength limit, \citet{Barger:2000p2144} obtained a formula describing the relation between the redshift and the submillimeter/radio flux ratio, $z = 0.98(S_{850~\mu{\rm m}}/S_{\rm 1.4~GHz})^{0.26} -1$. This implies z = 0.89 -- 0.99 for L3, which matches well to the measured spectroscopic redshift of $z=1.06$. Generally, AGN contamination could be one of the reasons responsible for the slight underestimation \citep{Carilli:1999p6658,Wang:2007p6971}. In this particular case, however, an inaccurate assumed dust temperature and radio flux contamination from the extended jet structure could also bias the estimation downward.

\subsection{A2390-1}

\citet{Cowie:2002p2075} discovered the bright SMG A2390-1 in their SCUBA survey, and its 850 $\mu$m flux is 7.57 $\pm$ 0.93 mJy. \citet{Metcalfe:2003p9536} performed a deep survey on A2390 using the ISOCAM on ESA's Infrared Space Observatory (ISO) and found a $\sim$ 3.7 $\sigma$ detection at 15 $\mu$m toward A2390-1. Our SMA data confirm the SCUBA detection with a flux of 5.55 $\pm$ 0.92 mJy obtained from the point-source fitting routine IMFIT with a box size of 10$''$.  The fitted position from IMFIT is $\alpha$(J2000.0) = 21$^h$53$^m$33.33$^s$, $\delta$(J2000.0) = 17$^\circ$42$^\prime$50.05$^{\prime\prime}$ with an error of 0$\farcs$7. Note that we only use the last two tracks of data to determine the source position because of high phase noise in the first two tracks of data.

There is a point source-like radio counterpart of A2390-1 in our latest A2390 VLA image (detailed information in Section 3.3). We compute the peak position and the flux by fitting a two dimensional gaussian with fixed major axis, minor axis, and phase angle obtained from the VLA synthesized beam. The radio position is at $\alpha$(J2000.0) = 21$^h$53$^m$33.31$^s$, $\delta$(J2000.0) = 17$^\circ$42$^\prime$50.33$^{\prime\prime}$ with 0$\farcs$03 error, which agrees with our SMA position. We adopt the position of A2390-1 measured from the VLA image, since it has a better accuracy. We use the VLA position and LENSTOOL to determine an amplification of 1.8 for A2390-1, which agrees well with the SCUBA measurement from \citet{Cowie:2002p2075}.

The fitting algorithm also gives a radio flux of 111.5 $\pm$ 6.5 $\mu$Jy (Table \ref{s}), which allows us to obtain the redshift of A2390-1, from the submillimeter/radio flux ratio technique. We find $z=1.8\pm0.2$. The infrared luminosity of A2390-1 that we estimate from the SED model described in Section 3.1 is 2.9 $\times$ 10$^{12}$~L$_\odot$. This shows that A2390-1 is also a ULIRG.

\subsection{A2390-3}
\begin{figure*}[t]
 \begin{center}
    \leavevmode
      \includegraphics[width=1 \textwidth] {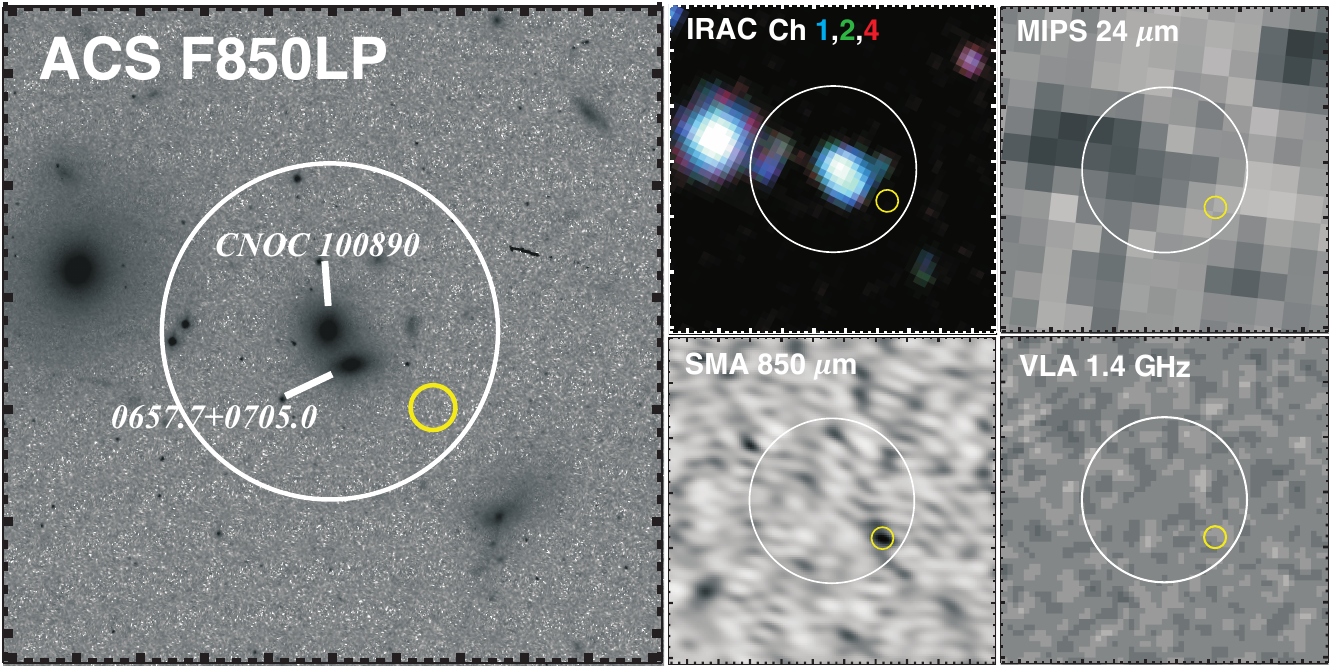}
       \caption{Multiwavelength images of A2390-3. North is up and East is to the left. The size of each image is 30$''$ $\times$ 30$''$. The big white circle in each image is centered at the position given in Table \ref{source} and represents the SCUBA beam size (15$''$ $\times$ 15$''$). The SMA position of A2390-3a is labeled in each image with a $2\farcs0$ diameter yellow circle. The gray scale images have inverse scales. Three of the IRAC channels (1, 2, and 4), corresponding to 3.6~$\mu$m, 4.5~$\mu$m, and 8.0~$\mu$m, are presented in the combined IRAC image with the color codes labeled.}
     \label{a2390_2}
  \end{center}
\end{figure*}

A2390-3 was first identified by \citet{Cowie:2002p2075} in their SCUBA map with a total flux of 3.24 $\pm$ 0.78~mJy. With seven times higher resolution, our SMA map detects the submillimeter object and appears to resolve it into two sources (Figure \ref{a2390sma}), which we refer to as A2390-3a and A2390-3b. The simultaneous point-source fitting routine IMFIT with a box size of 10$''$ gives fluxes of 3.15~mJy and 1.92~mJy, respectively. The r.m.s of the residual image is 0.58~mJy beam$^{-1}$, whereas the primary beam corrected noise is 0.63~mJy for A2390-3a and 0.60~mJy for A2390-3b. The fitted positions are $\alpha$(J2000.0) = 21$^h$53$^m$35.141$^s$, $\delta$(J2000.0) = 17$^\circ$41$^\prime$05.9$^{\prime\prime}$ with a $0\farcs4$ error for A2390-3a and $\alpha$(J2000.0) = 21$^h$53$^m$35.21$^s$, $\delta$(J2000.0) = 17$^\circ$41$^\prime$07.5$^{\prime\prime}$ with a $0\farcs6$ error for A2390-3b (Table \ref{s}). The total flux of 5.07 $\pm$ 0.87~mJy is consistent with the measured SCUBA flux. The error of 0.87 mJy is obtained from error propagation.

We optimized the output image (Figure \ref{a2390sma}) by applying the robust weighting of \citet{Briggs:1995p8411} with a robust parameter of +1.0 for the sake of making a balance between resolution and signal-to-noise ratio (S/N). To test the robustness of the detections, we iterated the process of making the image with different weightings on the Fourier transformation and different box sizes for the CLEAN algorithm. Noise spikes would vary or even disappear with the iterations; however, the signals from A2390-3a and A2390-3b remain, and the detections are always significant ($>3~\sigma$ for each component).

The amplification factor obtained by \citet{Cowie:2002p2075} on this source varies from 0.6 to 52 due to the positional uncertainty and redshift indeterminacy. However, with our arcsecond level precision, we can now obtain more accurate amplifications using LENSTOOL, which models the effects of the gravitational lensing by taking three dimensional mass distributions within the cluster into account \citep{Kneib:1996p3751}. 

We took the fitted positions and assumed that A2390-3a and A2390-3b are both point sources with a 1$''$ radius size on the image plane. LENSTOOL calculates the positions, semi-major axes, semi-minor axes, and inclination angles of the sources on the source plane. The amplification is then obtained by taking the inverse of the product of the semi-major and semi-minor axes (Table \ref{s}). We assume the source plane is located at $z = 5$ for the LENSTOOL calculation.  The amplifications of sources lying beyond $z = 1$ with modest amplifications have little dependence on redshift \citet{Blain:1999p7279}. We find an amplification of 2.3 for A2390-3a and 2.9 for A2390-3b.

We also present archival images from the \textit{Hubble Space Telescope (HST)} Advanced Camera for Surveys (ACS) \textit{z}-band (F850LP), the \textit{Spitzer} Infrared Array Camera (IRAC) 3.6~$\mu$m, 4.5~$\mu$m, and 8.0~$\mu$m bands, and the Multiband Imaging Photometer for \textit{Spitzer} (MIPS) 24~$\mu$m band (Figure \ref{a2390_2}). We present radio maps from VLA data obtained on October 20, 2008 at 1.4~GHz (Wang et al., in preparation), as well. With the most extended A configuration, the 1.4~GHz synthesized beam has a size of 1$\farcs4 \times 1\farcs4$, and the 1~$\sigma$ noise level is 6.5~$\mu$Jy beam$^{-1}$. Two cluster members lie within the SCUBA beam: CNOC 100890 was discovered by \citet{Yee:1996p7009} from the CNOC cluster redshift survey using the CFHT, while 0657.7+0705.0 was reported from a strong H$\alpha$ emission line search \citep{Balogh:2000p7036}. A strong amplification toward this region could be caused by these two galaxies.  

Interestingly, Figure~\ref{a2390_2} shows that there is no apparent detection corresponding to the SMA position (the SMA position of A2390-3a is shown) at other wavelengths. This evidence implies that both sources must be heavily embedded in dusty envelopes and/or at very high redshifts. For each source, the 1.4~GHz flux was measured from a given circular aperture which covers most of its submillimeter flux (Table \ref{s}), yet it is still a non-detection. We assigned zero to the VLA flux and adopted the 1~$\sigma$ errors obtained from Monte Carlo simulations as the upper limits on the radio fluxes, which are 5.2~$\mu$Jy and 3.3~$\mu$Jy for A2390-3a and A2390-3b, respectively. 

\begin{table*}
 \begin{center}
 \caption{Detected sources in the SMA observations}
 \scalebox{0.95}{
 \begin{tabular}{lccccccccc}
 \hline
 \hline
  Source & R.A. (2000) & Decl. (2000)  & S$_{850\mu m}$  & S$_{1.4 GHz}$ & z$^{a}$  & R.A. Offset  &  Dec. Offset & Amplification & $L_{IR}$ \\
  & (h m s) & (d m s) & (mJy) & ($\mu$Jy) &  & (arcsec) & (arcsec) & & ($L_\odot$) \\
 \hline
 A370-2 & 02 39 56.55 & -1 34 26.5 & 7.95 $\pm$ 0.60  & 573 $\pm$ 11 & 0.89 -- 0.99 & -1.2  &  0.5 & 2.3 & 2.5 $\times$ 10$^{12}$ \\
 A2390-1 & 21 53 33.31 & 17 42 50.3 & 5.55 $\pm$ 0.93 & 111 $\pm$ 7 & 1.60 -- 1.99 & 0.0$^c$ & 1.0$^c$ & 1.8 & 2.9 $\times$ 10$^{12}$  \\
 A2390-3 & ... & ... & 5.07 $\pm$ 0.87 & -4.0 $\pm$ 6.2 & $>$ 6.2 & ...  &  ...  & 2.5$^{b}$ & $>$  1.7 $\times$ 10$^{12}$\\
 A2390-3a & 21 53 35.14 & 17 41 05.9 & 3.15 $\pm$ 0.63 &-4.5 $\pm$ 5.2  & $>$ 5.2 & -4.8  &  -3.4 & 2.3 & $>$ 1.1 $\times$ 10$^{12}$ \\
 A2390-3b & 21 53 35.21 & 17 41 07.5 & 1.92 $\pm$ 0.60 & 0.5 $\pm$ 3.3 & $>$ 4.7 & -3.8 & -1.9 & 2.9 & $>$ 4.6 $\times$ 10$^{11}$\\ 
 \hline
  &&&&&&\\
 \multicolumn{10}{l}{$^{a}$ Estimated using the submillimeter to radio flux ratio \citep{Barger:2000p2144}.} \\
 \multicolumn{10}{l}{$^{b}$ Estimated under the assumption of a point source located in the middle of A2390-3a and A2390-3b on the source plane.} \\
 \multicolumn{10}{l}{$^{c}$ The SMA pointing position is different than the one reported in \citet{Cowie:2002p2075}. Here we use the reference position from}\\\multicolumn{10}{l}{\citet{Cowie:2002p2075}}\\
 \end{tabular}
 }
 \label{s}
  \end{center}
\end{table*}

Note that we are aware of the possibility of contamination from the residual sidelobes of the strong radio-bright cD galaxy in A2390. However, with the help of the latest data reduction techniques provided in AIPS, we find that the residual sidelobes from the cD galaxy can be mostly removed, except for the area close to the cD galaxy and lying generally N-S and away from the source. A2390-3a/b are located far from that area. Also, in our Monte Carlo simulations we avoid regions affected by bright sources and conspicuous residual sidelobes. Thus, we believe that our estimate of the 1 $\sigma$ flux limit is robust.

With the absence of a detection at all wavelengths except the submillimeter, it is not possible to obtain optical or infrared spectroscopic redshifts or photometric redshifts. A natural option would be to use the submillimeter/radio flux ratio analysis. Since the equation in Section~3.1 is limited to lower redshift sources ($z < 3$), we adopt more general equations (Eqs. (2) and (4) in \citealt{Barger:2000p2144}). We find $z > 5.2$ for A2390-3a and $z > 4.7$ for A2390-3b. If confirmed with an identification of CO lines, they could be among the most distant SMGs known. However, the high submillimeter/radio flux ratios could also be caused by using an incorrect dust temperature. In this case, the non-detection in the mid-infrared is a tentative indication that the dust temperature is likely to be low. Also, SMGs tend to have lower dust temperatures than their local infrared counterparts \citep{Pope:2006p8076, Huynh:2007p8434, Hwang:2010p8724}. Any dust temperature lower than our adopted one would move the source even further (i.e., to higher redshifts). Thus, our estimation of the redshifts should be reasonable.

Having the lower limit redshift and amplification information for both sources in hand, we can also compute lower limits to the infrared luminosities (Table~\ref{s}). The star formation rates can be estimated if we assume starburst galaxies and a standard Salpeter initial mass function. In this work, we use the formula $\dot{M} = 1.7 \times 10^{-10}~L_{FIR}/L_{\odot}$ \citep{Kennicutt:1998p5718}, which gives 188 and 78~$M_\odot /$yr for A2390-3a and A2390-3b.

It is possible that these two sources are coincidently passing through the line-of-sight at different distances. Our estimates of the redshifts are rough because of the uncertainty in the radio-FIR correlation. LENSTOOL shows that the separation between the two sources goes from $1\farcs8$ on the image plane down to $1\farcs7$ on the source plane. More interestingly, if we assume a point source located in the middle of the two sources on the source plane with a 1$''$ radius morphology, then the gravitationally lensed image on the image plane appears elongated and covers both sources (green ellipse in Fig. \ref{a2390sma}).  This implies that it is also possible (perhaps more likely) that these two sources are the outcome of the distortion by gravitational lensing of a close pair or of one source broken into two. At this point we cannot be secure about the nature of this object, and more rigorous investigations, such as CO line searches, would be needed to proceed further. We have also tried treating the object as one source, and we present that information in Table~\ref{s}, as well. 

\section{Discussion}

\begin{figure}[h]
 \begin{center}
    \leavevmode
        \includegraphics[scale=0.4]{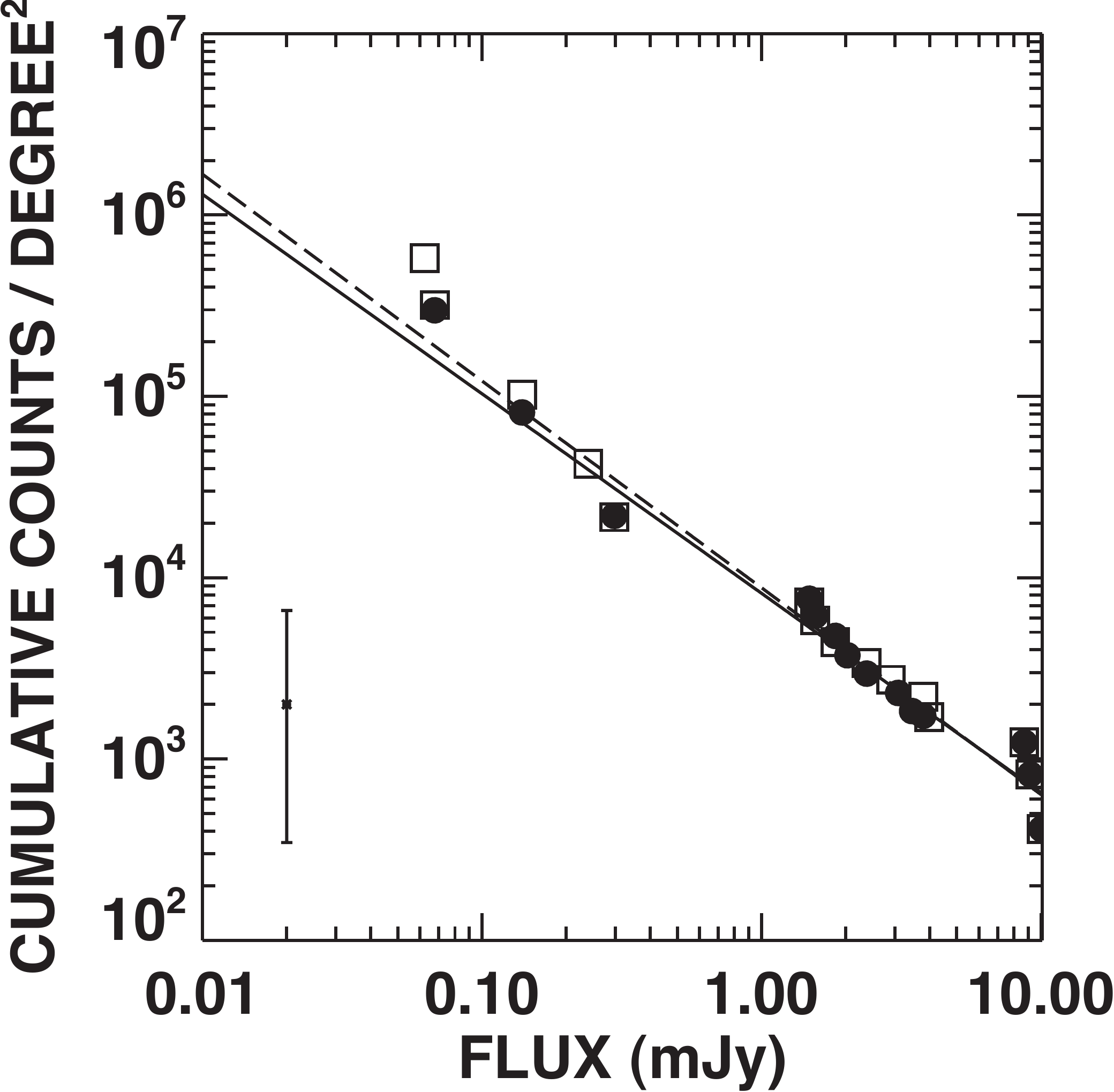}
       \caption{Updated cumulative 850~$\mu$m source counts (filled circles) using the method of \citet{Cowie:2002p2075} (open squares). The solid curve is an area-weighted maximum likelihood power-law fit to the updated data points over the range 0.1 to 4.0~mJy with a slope of $-1.10$. The dashed curve is the fit to the original data and has a slope of $-1.14$. The typical error bar shown in the bottom left corner assumes Poisson statistics.}
     \label{cnc}
  \end{center}
\end{figure}

Previous SMA studies have focused on very luminous SMGs \citep{Iono:2006p6985,Younger:2007p6982,Wang:2007p6971,Cowie:2009p6978}. Those bright SMGs have a total infrared luminosity in the range of 10$^{12}$ -- 10$^{13}$ $L_\odot$, depending on the redshift. While A370-2 and A2390-1 are also ULIRGs, compared to those giant galaxies, A2390-3b is much more typical, thanks to the gravitational lensing. Even if we assume that A2390-3b has a redshift of 6, the infrared luminosity still remains below 5 $\times$ 10$^{11}$ $L_\odot$. Moreover, our non-detection toward A2390-4 may also imply multiple faint sources. Given our detection limit (0.96~mJy for 3~$\sigma$), the assumed redshift ($z = 3$ obtained from the submillimeter/radio flux ratio), and the amplification (6.7), the total FIR luminosity would easily be of the order of typical galaxies ($\sim$ 1.4 $\times$ 10$^{11}$ $L_\odot$).

Giant SMGs tend to dominate the universal star formation history during the epoch z = 1 -- 3 \citep{Chapman:2005p5778,Wang:2006p2031}. The answer to the question of whether this situation continues to higher redshifts is critical for understanding the star formation history. Smaller galaxies like A2390-3b are expected to dominate the light in the very early Universe ($z\sim 5$ or higher), or else bottom-up clumping cold dark matter models of galaxy formation will be seriously challenged. Thus, obtaining the redshifts for A2390-3b and other submillimeter-faint sources, possibly through CO observations, is an important next step.

With our updated positional information, we have recomputed the 850~$\mu$m cumulative number counts shown in \citet{Cowie:2002p2075}. We used the data in \citet{Cowie:2002p2075} and updated the three sources in this work with newly obtained information. The results are shown in Figure \ref{cnc}. We adopt the method described in \citet{Cowie:2002p2075} to calculate the source areas and cumulative counts. We calculate the upper and lower error bars using Poisson statistics, and we use the table from \citet{Gehrels:1986p1344} to find the confidence limits on small numbers. We also perform an area-weighted maximum likelihood fit to the updated data over the range from 0.1 to 4.0~mJy, which gives the best fit equation of 
\begin{equation}
N(>S) = 8.2 \times 10^3 S^{-1.10}. \nonumber
\end{equation}
The slope of the fit is slightly shallower than the previous fit ($-1.14$) using the same clusters. Note that we assume no source in A2390-4 and take A2390-3 to be a single source. 

\begin{figure}[h]
 \begin{center}
    \leavevmode
        \includegraphics[scale=0.42]{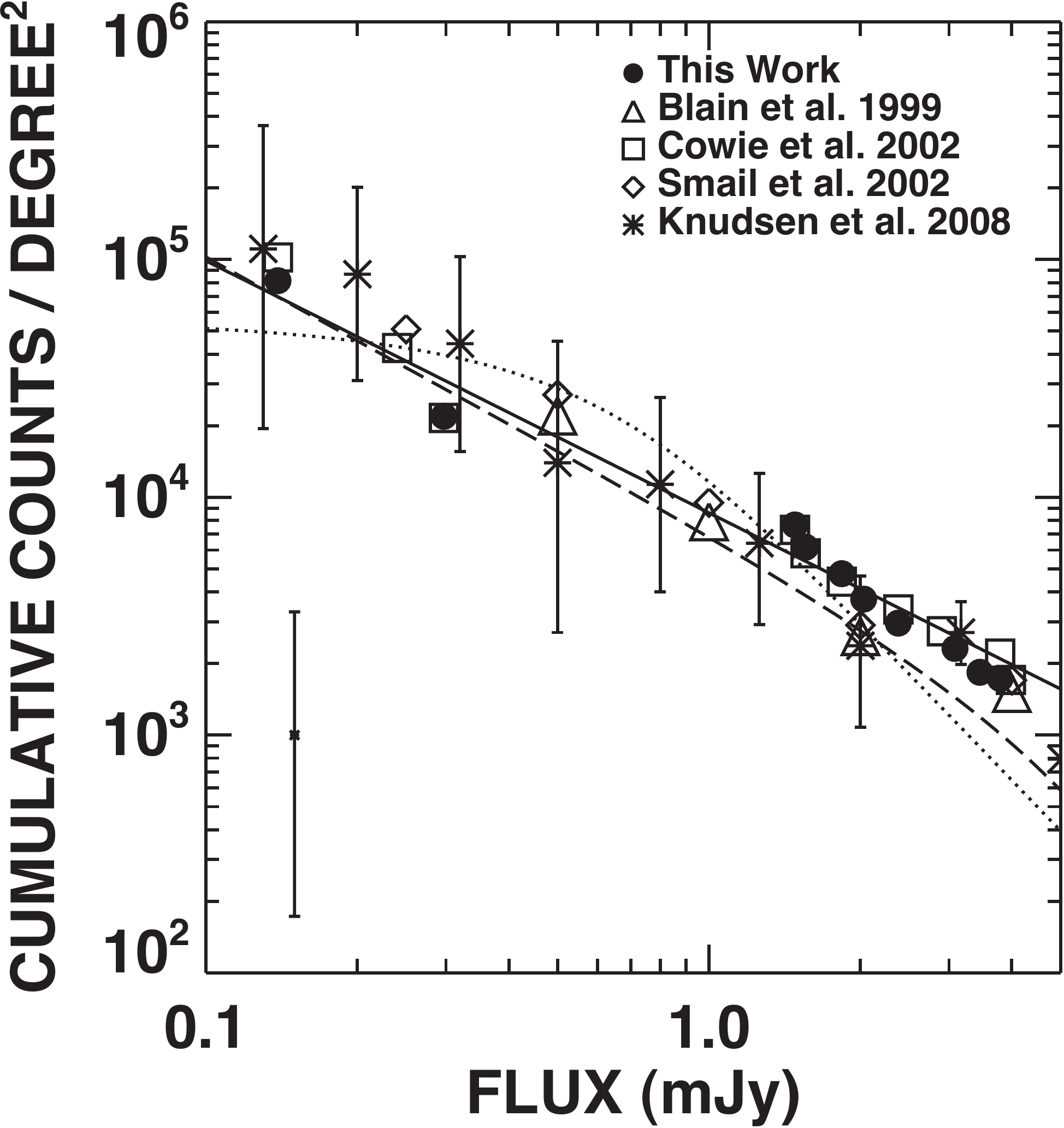}
       \caption{Cumulative 850~$\mu$m source counts from other lensing surveys compared with the updated source counts from this work. For clarity we only show the error bars on the \citet{Knudsen:2008p3824} data points. The typical 1~$\sigma$ error bar for our counts is shown in the bottom left corner. The solid curve is our power-law fit to the updated source counts with a slope of -1.10, the dashed curve is the double power-law fit from \citet{Knudsen:2008p3824}, and the dotted curve is the parametric fit from \citet{Barger:1999p6485}.}
       
     \label{cnc_com}
  \end{center}
\end{figure}

In Figure~\ref{cnc_com} we compare our updated results at the faint end (0.1~mJy $<$ $S_{850}$ $<$ 4~mJy) with previous lensing surveys \citep{Blain:1999p7279, Cowie:2002p2075,Smail:2002p6793,Knudsen:2008p3824}. We overplot various fitted curves: the solid line is the power-law fit in this work, the dashed line shows the double power-law fit described in \citet{Knudsen:2008p3824}, and the dotted curve shows an empirical fit to a blank field survey \citep{Barger:1999p6485} that was constrained by the 850~$\mu$m EBL measurements at lower fluxes. All three curves reasonably describe the cumulative counts, though Barger et al.'s model may be preferred as a better match if the trend of flattening continues to counts lower than 0.1~mJy. Our faint-end counts are slightly lower but still in a good agreement with the previous literature.

The changes in our derived counts emphasize the uncertainty introduced in the lensing analysis when the source positions are poorly determined. The reason for the change is that the faintest sources get a large boost from gravitational lensing (amplifications $>$ 10), and for these sources the variations in the amplifications due to positional errors can be an order of magnitude \citep{Cowie:2002p2075, Knudsen:2008p3824}. The small source areas at the faint fluxes then cause large changes. Using A2390-3 as an example, the accurate position from our results brings the amplification from 52 down to 2.5 assuming one source. In fact, our updated flux and amplification for A2390-3 is the main reason for the shallower fit.

\begin{figure}[h]
 \begin{center}
    \leavevmode
          \includegraphics[scale=0.43]{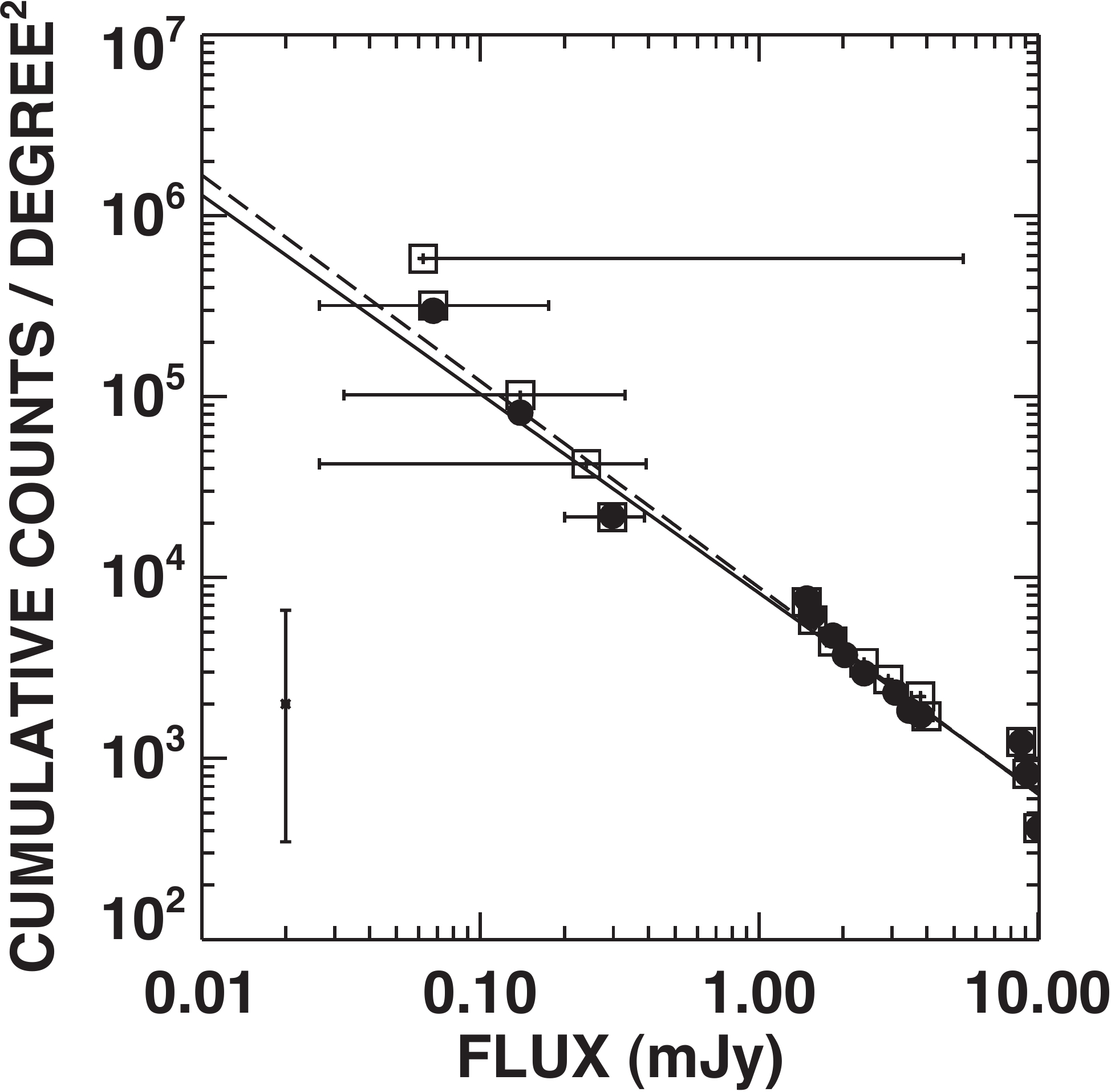}
       \caption{Same as Figure \ref{cnc}, but with horizontal error bars shown on the 850~$\mu$m fluxes of the sources in \citet{Cowie:2002p2075}, which are caused by amplification uncertainties.}
     \label{cnc_xerr}
  \end{center}
\end{figure}

In Figure~\ref{cnc_xerr} we show the \citet{Cowie:2002p2075} data points with error bars on the 850~$\mu$m fluxes, which are due to the uncertainties in the amplifications (i.e., resulting from the positional uncertainties from the SCUBA measurements).  Large flux errors on the faint-end sources clearly demonstrate the problems of using measurements from single-dish telescopes. The fact that our shallower power-law fit results from our accurate determination of the amplification of A2390-3 implies that accurate positions of SMGs are not only critical for finding the correct counterparts in order to determine the nature of individual SMGs, but also for measuring the overall properties of the submillimeter sources. 

Moreover, the multiplicity of SMGs has been implied from several indirect studies using 24 $\mu$m \citep{Pope:2006p8076}, 350 $\mu$m \citep{Kovacs:2010p9532}, and radio \citep{Ivison:2007p9495,Younger:2009p9502}, as well as CO interferometry \citep{Tacconi:2006p8449,Tacconi:2008p9334,Engel:2010p9470}. The most recent discovery by \citet{Wang:2011p9293} in the GOODS-N region that some SCUBA-detected SMGs split into multiple fainter sources in deep SMA imaging supports this result. This multiplicity of SMGs can potentially increase the number counts at the low luminosity end and may also fundamentally change the luminosity function of SMGs. Thus, future high-resolution observations using interferometers like the SMA or ALMA will be critical for obtaining a better understanding of the star formation history.

\section{Summary}
We have reported the results of our SMA observations of four SCUBA-detected sources, A370-2, A2390-1, A2390-3 and A2390-4. Although it had been suggested with indirect evidence that  the optical source L3 was the likely counterpart of A370-2, our direct detection of submillimeter emission from L3 unambiguously confirms this. A2390-1 is a ULIRG with a submillimeter flux of 5.55 $\pm$ 0.92~mJy and a redshift of 1.8 $\pm$ 0.2 computed from submillimeter/radio flux ratio analysis. We detected two lensed SMA sources toward A2390-3. The answer to the question of whether these two sources are physically related or the consequence of lensing distortion is still unclear. The fact that no counterparts are found in optical and infrared images indicates a heavily dust-embedded nature, and a lack of radio emission implies that both sources are located beyond $z = 4.7$. Our non-detection toward A2390-4 suggests either that there are multiple sources within the SCUBA beam or that the SCUBA detection may have been false. The accurate amplifications and fluxes from this work provide a shallower slope in the faint-end 850~$\mu$m cumulative number counts. Our results emphasize the need for high-resolution observations toward single-dish detected SMGs. 
 
\acknowledgments
We gratefully acknowledge support from NSF grants AST 0709356 (C.C.C., L.L.C.) and AST 0708793 (A.J.B.), National Science Council of Taiwan grant 
99-2112-M-001-012-MY3 (W.-H.W.), the University of Wisconsin Research Committee with funds granted by the Wisconsin Alumni Research Foundation (A.J.B.), and the David and Lucile Packard Foundation (A.J.B.).
\bibliography{bib}

\begin{thebibliography}{61}
\expandafter\ifx\csname natexlab\endcsname\relax\def\natexlab#1{#1}\fi

\bibitem[{{Alexander} {et~al.}(2005){Alexander}, {Bauer}, {Chapman}, {Smail},
  {Blain}, {Brandt}, \& {Ivison}}]{Alexander:2005p6453}
{Alexander}, D.~M., {et~al.} 2005, \apj, 632, 736

\bibitem[{{Aravena} {et~al.}(2010){Aravena}, {Younger}, {Fazio}, {Gurwell},
  {Espada}, {Bertoldi}, {Capak}, \& {Wilner}}]{Aravena:2010p8370}
{Aravena}, M., {et~al.} 2010, \apjl, 719, L15

\bibitem[{{Balogh} \& {Morris}(2000)}]{Balogh:2000p7036}
{Balogh}, M.~L., \& {Morris}, S.~L. 2000, \mnras, 318, 703

\bibitem[{{Barger} {et~al.}(2000){Barger}, {Cowie}, \&
  {Richards}}]{Barger:2000p2144}
{Barger}, A.~J., {Cowie}, L.~L., \& {Richards}, E.~A. 2000, \aj, 119, 2092

\bibitem[{{Barger} {et~al.}(1999{\natexlab{a}}){Barger}, {Cowie}, \&
  {Sanders}}]{Barger:1999p6485}
{Barger}, A.~J., {Cowie}, L.~L., \& {Sanders}, D.~B. 1999{\natexlab{a}}, \apjl,
  518, L5

\bibitem[{{Barger} {et~al.}(1999{\natexlab{b}}){Barger}, {Cowie}, {Smail},
  {Ivison}, {Blain}, \& {Kneib}}]{Barger:1999p6801}
{Barger}, A.~J., {et~al.} 1999{\natexlab{b}}, \aj, 117, 2656

\bibitem[{{Blain} {et~al.}(1999){Blain}, {Kneib}, {Ivison}, \&
  {Smail}}]{Blain:1999p7279}
{Blain}, A.~W., {et~al.} 1999, \apjl, 512, L87

\bibitem[{{Blain} {et~al.}(2002){Blain}, {Smail}, {Ivison}, {Kneib}, \&
  {Frayer}}]{Blain:2002p8120}
---. 2002, \physrep, 369, 111

\bibitem[{{Borys} {et~al.}(2003){Borys}, {Chapman}, {Halpern}, \&
  {Scott}}]{Borys:2003p6612}
{Borys}, C., {et~al.} 2003, \mnras, 344, 385

\bibitem[{{Briggs}(1995)}]{Briggs:1995p8411}
{Briggs}, D.~S. 1995, in Bulletin of the American Astronomical Society,
  Vol.~27, Bulletin of the American Astronomical Society, 1444--+

\bibitem[{{Capak} {et~al.}(2008){Capak}, {Carilli}, {Lee}, {Aldcroft},
  {Aussel}, {Schinnerer}, {Wilson}, {Yun}, {Blain}, {Giavalisco}, {Ilbert},
  {Kartaltepe}, {Lee}, {McCracken}, {Mobasher}, {Salvato}, {Sasaki}, {Scott},
  {Sheth}, {Shioya}, {Thompson}, {Elvis}, {Sanders}, {Scoville}, \&
  {Tanaguchi}}]{Capak:2008p8406}
{Capak}, P., {et~al.} 2008, \apjl, 681, L53

\bibitem[{{Carilli} \& {Yun}(1999)}]{Carilli:1999p6658}
{Carilli}, C.~L., \& {Yun}, M.~S. 1999, \apjl, 513, L13

\bibitem[{{Chapman} {et~al.}(2005){Chapman}, {Blain}, {Smail}, \&
  {Ivison}}]{Chapman:2005p5778}
{Chapman}, S.~C., {et~al.} 2005, \apj, 622, 772

\bibitem[{{Condon}(1992)}]{Condon:1992p6652}
{Condon}, J.~J. 1992, \araa, 30, 575

\bibitem[{{Coppin} {et~al.}(2006){Coppin}, {Chapin}, {Mortier}, {Scott},
  {Borys}, {Dunlop}, {Halpern}, {Hughes}, {Pope}, {Scott}, {Serjeant}, {Wagg},
  {Alexander}, {Almaini}, {Aretxaga}, {Babbedge}, {Best}, {Blain}, {Chapman},
  {Clements}, {Crawford}, {Dunne}, {Eales}, {Edge}, {Farrah}, {Gazta{\~n}aga},
  {Gear}, {Granato}, {Greve}, {Fox}, {Ivison}, {Jarvis}, {Jenness}, {Lacey},
  {Lepage}, {Mann}, {Marsden}, {Martinez-Sansigre}, {Oliver}, {Page},
  {Peacock}, {Pearson}, {Percival}, {Priddey}, {Rawlings}, {Rowan-Robinson},
  {Savage}, {Seigar}, {Sekiguchi}, {Silva}, {Simpson}, {Smail}, {Stevens},
  {Takagi}, {Vaccari}, {van Kampen}, \& {Willott}}]{Coppin:2006p9123}
{Coppin}, K., {et~al.} 2006, \mnras, 372, 1621

\bibitem[{{Coppin} {et~al.}(2009){Coppin}, {Smail}, {Alexander}, {Weiss},
  {Walter}, {Swinbank}, {Greve}, {Kovacs}, {De Breuck}, {Dickinson}, {Ibar},
  {Ivison}, {Reddy}, {Spinrad}, {Stern}, {Brandt}, {Chapman}, {Dannerbauer},
  {van Dokkum}, {Dunlop}, {Frayer}, {Gawiser}, {Geach}, {Huynh}, {Knudsen},
  {Koekemoer}, {Lehmer}, {Menten}, {Papovich}, {Rix}, {Schinnerer}, {Wardlow},
  \& {van der Werf}}]{Coppin:2009p8366}
{Coppin}, K.~E.~K., {et~al.} 2009, \mnras, 395, 1905

\bibitem[{{Coppin} {et~al.}(2010){Coppin}, {Chapman}, {Smail}, {Swinbank},
  {Walter}, {Wardlow}, {Weiss}, {Alexander}, {Brandt}, {Dannerbauer}, {De
  Breuck}, {Dickinson}, {Dunlop}, {Edge}, {Emonts}, {Greve}, {Huynh}, {Ivison},
  {Knudsen}, {Menten}, {Schinnerer}, \& {van der Werf}}]{Coppin:2010p8367}
---. 2010, \mnras, 407, L103

\bibitem[{{Cowie} {et~al.}(2002){Cowie}, {Barger}, \&
  {Kneib}}]{Cowie:2002p2075}
{Cowie}, L.~L., {Barger}, A.~J., \& {Kneib}, J. 2002, \aj, 123, 2197

\bibitem[{{Cowie} {et~al.}(2009){Cowie}, {Barger}, {Wang}, \&
  {Williams}}]{Cowie:2009p6978}
{Cowie}, L.~L., {et~al.} 2009, \apjl, 697, L122

\bibitem[{{Daddi} {et~al.}(2009{\natexlab{a}}){Daddi}, {Dannerbauer}, {Krips},
  {Walter}, {Dickinson}, {Elbaz}, \& {Morrison}}]{Daddi:2009p9297}
{Daddi}, E., {et~al.} 2009{\natexlab{a}}, \apjl, 695, L176

\bibitem[{{Daddi} {et~al.}(2009{\natexlab{b}}){Daddi}, {Dannerbauer}, {Stern},
  {Dickinson}, {Morrison}, {Elbaz}, {Giavalisco}, {Mancini}, {Pope}, \&
  {Spinrad}}]{Daddi:2009p9295}
---. 2009{\natexlab{b}}, \apj, 694, 1517

\bibitem[{{Eales} {et~al.}(2003){Eales}, {Bertoldi}, {Ivison}, {Carilli},
  {Dunne}, \& {Owen}}]{Eales:2003p6489}
{Eales}, S., {et~al.} 2003, \mnras, 344, 169

\bibitem[{{Engel} {et~al.}(2010){Engel}, {Tacconi}, {Davies}, {Neri}, {Smail},
  {Chapman}, {Genzel}, {Cox}, {Greve}, {Ivison}, {Blain}, {Bertoldi}, \&
  {Omont}}]{Engel:2010p9470}
{Engel}, H., {et~al.} 2010, \apj, 724, 233

\bibitem[{{Gehrels}(1986)}]{Gehrels:1986p1344}
{Gehrels}, N. 1986, \apj, 303, 336

\bibitem[{{Greve} {et~al.}(2004){Greve}, {Ivison}, {Bertoldi}, {Stevens},
  {Dunlop}, {Lutz}, \& {Carilli}}]{Greve:2004p6618}
{Greve}, T.~R., {et~al.} 2004, \mnras, 354, 779

\bibitem[{{Greve} {et~al.}(2005){Greve}, {Bertoldi}, {Smail}, {Neri},
  {Chapman}, {Blain}, {Ivison}, {Genzel}, {Omont}, {Cox}, {Tacconi}, \&
  {Kneib}}]{Greve:2005p6788}
---. 2005, \mnras, 359, 1165

\bibitem[{{Ho} {et~al.}(2004){Ho}, {Moran}, \& {Lo}}]{Ho:2004p8376}
{Ho}, P.~T.~P., {Moran}, J.~M., \& {Lo}, K.~Y. 2004, \apjl, 616, L1

\bibitem[{{Huynh} {et~al.}(2007){Huynh}, {Pope}, {Frayer}, \&
  {Scott}}]{Huynh:2007p8434}
{Huynh}, M.~T., {et~al.} 2007, \apj, 659, 305

\bibitem[{{Hwang} {et~al.}(2010){Hwang}, {Elbaz}, {Magdis}, {Daddi},
  {Symeonidis}, {Altieri}, {Amblard}, {Andreani}, {Arumugam}, {Auld}, {Aussel},
  {Babbedge}, {Berta}, {Blain}, {Bock}, {Bongiovanni}, {Boselli}, {Buat},
  {Burgarella}, {Castro-Rodr{\'{\i}}guez}, {Cava}, {Cepa}, {Chanial}, {Chapin},
  {Chary}, {Cimatti}, {Clements}, {Conley}, {Conversi}, {Cooray},
  {Dannerbauer}, {Dickinson}, {Dominguez}, {Dowell}, {Dunlop}, {Dwek}, {Eales},
  {Farrah}, {Schreiber}, {Fox}, {Franceschini}, {Gear}, {Genzel}, {Glenn},
  {Griffin}, {Gruppioni}, {Halpern}, {Hatziminaoglou}, {Ibar}, {Isaak},
  {Ivison}, {Jeong}, {Lagache}, {Le Borgne}, {Le Floc'h}, {Lee}, {Lee}, {Lee},
  {Levenson}, {Lu}, {Lutz}, {Madden}, {Maffei}, {Magnelli}, {Mainetti},
  {Maiolino}, {Marchetti}, {Mortier}, {Nguyen}, {Nordon}, {O'Halloran},
  {Okumura}, {Oliver}, {Omont}, {Page}, {Panuzzo}, {Papageorgiou}, {Pearson},
  {P{\'e}rez-Fournon}, {Garc{\'{\i}}a}, {Poglitsch}, {Pohlen}, {Popesso},
  {Pozzi}, {Rawlings}, {Rigopoulou}, {Riguccini}, {Rizzo}, {Rodighiero},
  {Roseboom}, {Rowan-Robinson}, {Saintonge}, {Portal}, {Santini}, {Sauvage},
  {Schulz}, {Scott}, {Seymour}, {Shao}, {Shupe}, {Smith}, {Stevens}, {Sturm},
  {Tacconi}, {Trichas}, {Tugwell}, {Vaccari}, {Valtchanov}, {Vieira},
  {Vigroux}, {Wang}, {Ward}, {Wright}, {Xu}, \& {Zemcov}}]{Hwang:2010p8724}
{Hwang}, H.~S., {et~al.} 2010, \mnras, 409, 75

\bibitem[{{Iono} {et~al.}(2006){Iono}, {Peck}, {Pope}, {Borys}, {Scott},
  {Wilner}, {Gurwell}, {Ho}, {Yun}, {Matsushita}, {Petitpas}, {Dunlop},
  {Elvis}, {Blain}, \& {Le Floc'h}}]{Iono:2006p6985}
{Iono}, D., {et~al.} 2006, \apjl, 640, L1

\bibitem[{{Ivison} {et~al.}(2007){Ivison}, {Greve}, {Dunlop}, {Peacock},
  {Egami}, {Smail}, {Ibar}, {van Kampen}, {Aretxaga}, {Babbedge}, {Biggs},
  {Blain}, {Chapman}, {Clements}, {Coppin}, {Farrah}, {Halpern}, {Hughes},
  {Jarvis}, {Jenness}, {Jones}, {Mortier}, {Oliver}, {Papovich},
  {P{\'e}rez-Gonz{\'a}lez}, {Pope}, {Rawlings}, {Rieke}, {Rowan-Robinson},
  {Savage}, {Scott}, {Seigar}, {Serjeant}, {Simpson}, {Stevens}, {Vaccari},
  {Wagg}, \& {Willott}}]{Ivison:2007p9495}
{Ivison}, R.~J., {et~al.} 2007, \mnras, 380, 199

\bibitem[{{Kennicutt}(1998)}]{Kennicutt:1998p5718}
{Kennicutt}, Jr., R.~C. 1998, \apj, 498, 541

\bibitem[{{Kneib} {et~al.}(1996){Kneib}, {Ellis}, {Smail}, {Couch}, \&
  {Sharples}}]{Kneib:1996p3751}
{Kneib}, J., {et~al.} 1996, \apj, 471, 643

\bibitem[{{Knudsen} {et~al.}(2008){Knudsen}, {van der Werf}, \&
  {Kneib}}]{Knudsen:2008p3824}
{Knudsen}, K.~K., {van der Werf}, P.~P., \& {Kneib}, J. 2008, \mnras, 384, 1611

\bibitem[{{Komatsu} {et~al.}(2009){Komatsu}, {Dunkley}, {Nolta}, {Bennett},
  {Gold}, {Hinshaw}, {Jarosik}, {Larson}, {Limon}, {Page}, {Spergel},
  {Halpern}, {Hill}, {Kogut}, {Meyer}, {Tucker}, {Weiland}, {Wollack}, \&
  {Wright}}]{Komatsu:2009p8122}
{Komatsu}, E., {et~al.} 2009, \apjs, 180, 330

\bibitem[{{Kov{\'a}cs} {et~al.}(2010){Kov{\'a}cs}, {Omont}, {Beelen},
  {Lonsdale}, {Polletta}, {Fiolet}, {Greve}, {Borys}, {Cox}, {De Breuck},
  {Dole}, {Dowell}, {Farrah}, {Lagache}, {Menten}, {Bell}, \&
  {Owen}}]{Kovacs:2010p9532}
{Kov{\'a}cs}, A., {et~al.} 2010, \apj, 717, 29

\bibitem[{{Mellier} {et~al.}(1988){Mellier}, {Soucail}, {Fort}, \&
  {Mathez}}]{Mellier:1988p6845}
{Mellier}, Y., {et~al.} 1988, \aap, 199, 13

\bibitem[{{Metcalfe} {et~al.}(2003){Metcalfe}, {Kneib}, {McBreen}, {Altieri},
  {Biviano}, {Delaney}, {Elbaz}, {Kessler}, {Leech}, {Okumura}, {Ott},
  {Perez-Martinez}, {Sanchez-Fernandez}, \& {Schulz}}]{Metcalfe:2003p9536}
{Metcalfe}, L., {et~al.} 2003, \aap, 407, 791

\bibitem[{{Morrison} {et~al.}(2010){Morrison}, {Owen}, {Dickinson}, {Ivison},
  \& {Ibar}}]{Morrison:2010p7451}
{Morrison}, G.~E., {et~al.} 2010, \apjs, 188, 178

\bibitem[{{Pope} {et~al.}(2006){Pope}, {Scott}, {Dickinson}, {Chary},
  {Morrison}, {Borys}, {Sajina}, {Alexander}, {Daddi}, {Frayer}, {MacDonald},
  \& {Stern}}]{Pope:2006p8076}
{Pope}, A., {et~al.} 2006, \mnras, 370, 1185

\bibitem[{{Riechers} {et~al.}(2010){Riechers}, {Capak}, {Carilli}, {Cox},
  {Neri}, {Scoville}, {Schinnerer}, {Bertoldi}, \& {Yan}}]{Riechers:2010p9307}
{Riechers}, D.~A., {et~al.} 2010, \apjl, 720, L131

\bibitem[{{Sanders} \& {Mirabel}(1996)}]{Sanders:1996p6419}
{Sanders}, D.~B., \& {Mirabel}, I.~F. 1996, \araa, 34, 749

\bibitem[{{Sault} {et~al.}(1995){Sault}, {Teuben}, \&
  {Wright}}]{Sault:1995p6974}
{Sault}, R.~J., {Teuben}, P.~J., \& {Wright}, M.~C.~H. 1995, in Astronomical
  Society of the Pacific Conference Series, Vol.~77, Astronomical Data Analysis
  Software and Systems IV, ed. {R.~A.~Shaw, H.~E.~Payne, \& J.~J.~E.~Hayes},
  433--+

\bibitem[{{Schinnerer} {et~al.}(2008){Schinnerer}, {Carilli}, {Capak},
  {Martinez-Sansigre}, {Scoville}, {Smol{\v c}i{\'c}}, {Taniguchi}, {Yun},
  {Bertoldi}, {Le Fevre}, \& {de Ravel}}]{Schinnerer:2008p9300}
{Schinnerer}, E., {et~al.} 2008, \apjl, 689, L5

\bibitem[{{Scott} {et~al.}(2002){Scott}, {Fox}, {Dunlop}, {Serjeant},
  {Peacock}, {Ivison}, {Oliver}, {Mann}, {Lawrence}, {Efstathiou},
  {Rowan-Robinson}, {Hughes}, {Archibald}, {Blain}, \&
  {Longair}}]{Scott:2002p6539}
{Scott}, S.~E., {et~al.} 2002, \mnras, 331, 817

\bibitem[{{Smail} {et~al.}(2002){Smail}, {Ivison}, {Blain}, \&
  {Kneib}}]{Smail:2002p6793}
{Smail}, I., {et~al.} 2002, \mnras, 331, 495

\bibitem[{{Smail} {et~al.}(2000){Smail}, {Ivison}, {Owen}, {Blain}, \&
  {Kneib}}]{Smail:2000p6377}
---. 2000, \apj, 528, 612

\bibitem[{{Soucail} {et~al.}(1999){Soucail}, {Kneib}, {B{\'e}zecourt},
  {Metcalfe}, {Altieri}, \& {Le Borgne}}]{Soucail:1999p6372}
{Soucail}, G., {et~al.} 1999, \aap, 343, L70

\bibitem[{{Tacconi} {et~al.}(2006){Tacconi}, {Neri}, {Chapman}, {Genzel},
  {Smail}, {Ivison}, {Bertoldi}, {Blain}, {Cox}, {Greve}, \&
  {Omont}}]{Tacconi:2006p8449}
{Tacconi}, L.~J., {et~al.} 2006, \apj, 640, 228

\bibitem[{{Tacconi} {et~al.}(2008){Tacconi}, {Genzel}, {Smail}, {Neri},
  {Chapman}, {Ivison}, {Blain}, {Cox}, {Omont}, {Bertoldi}, {Greve},
  {F{\"o}rster Schreiber}, {Genel}, {Lutz}, {Swinbank}, {Shapley}, {Erb},
  {Cimatti}, {Daddi}, \& {Baker}}]{Tacconi:2008p9334}
---. 2008, \apj, 680, 246

\bibitem[{{Taniguchi} \& {Murayama}(2001)}]{Taniguchi:2001p6381}
{Taniguchi}, Y., \& {Murayama}, T. 2001, \apjl, 547, L13

\bibitem[{{Wang} {et~al.}(2004){Wang}, {Cowie}, \& {Barger}}]{Wang:2004p2270}
{Wang}, W., {Cowie}, L.~L., \& {Barger}, A.~J. 2004, \apj, 613, 655

\bibitem[{{Wang} {et~al.}(2006){Wang}, {Cowie}, \& {Barger}}]{Wang:2006p2031}
---. 2006, \apj, 647, 74

\bibitem[{{Wang} {et~al.}(2011){Wang}, {Cowie}, {Barger}, \&
  {Williams}}]{Wang:2011p9293}
{Wang}, W., {et~al.} 2011, \apjl, 726, L18+

\bibitem[{{Wang} {et~al.}(2007){Wang}, {Cowie}, {van Saders}, {Barger}, \&
  {Williams}}]{Wang:2007p6971}
---. 2007, \apjl, 670, L89

\bibitem[{{Webb} {et~al.}(2003){Webb}, {Eales}, {Lilly}, {Clements}, {Dunne},
  {Gear}, {Ivison}, {Flores}, \& {Yun}}]{Webb:2003p6591}
{Webb}, T.~M., {et~al.} 2003, \apj, 587, 41

\bibitem[{{Wu} {et~al.}(2009){Wu}, {Vanden Bout}, {Evans}, \&
  {Dunham}}]{Wu:2009p6792}
{Wu}, J., {et~al.} 2009, \apj, 707, 988

\bibitem[{{Yee} {et~al.}(1996){Yee}, {Ellingson}, {Abraham}, {Gravel},
  {Carlberg}, {Smecker-Hane}, {Schade}, \& {Rigler}}]{Yee:1996p7009}
{Yee}, H.~K.~C., {et~al.} 1996, \apjs, 102, 289

\bibitem[{{Younger} {et~al.}(2007){Younger}, {Fazio}, {Huang}, {Yun}, {Wilson},
  {Ashby}, {Gurwell}, {Lai}, {Peck}, {Petitpas}, {Wilner}, {Iono}, {Kohno},
  {Kawabe}, {Hughes}, {Aretxaga}, {Webb}, {Mart{\'{\i}}nez-Sansigre}, {Kim},
  {Scott}, {Austermann}, {Perera}, {Lowenthal}, {Schinnerer}, \& {Smol{\v
  c}i{\'c}}}]{Younger:2007p6982}
{Younger}, J.~D., {et~al.} 2007, \apj, 671, 1531

\bibitem[{{Younger} {et~al.}(2008){Younger}, {Dunlop}, {Peck}, {Ivison},
  {Biggs}, {Chapin}, {Clements}, {Dye}, {Greve}, {Hughes}, {Iono}, {Smail},
  {Krips}, {Petitpas}, {Wilner}, {Schael}, \& {Wilson}}]{Younger:2008p8372}
---. 2008, \mnras, 387, 707

\bibitem[{{Younger} {et~al.}(2009){Younger}, {Fazio}, {Huang}, {Yun}, {Wilson},
  {Ashby}, {Gurwell}, {Peck}, {Petitpas}, {Wilner}, {Hughes}, {Aretxaga},
  {Kim}, {Scott}, {Austermann}, {Perera}, \& {Lowenthal}}]{Younger:2009p9502}
---. 2009, \apj, 704, 803

\end{thebibliography}

\end{document}